\documentclass{article}
\usepackage{arxiv}

\usepackage[utf8]{inputenc} 
\usepackage[T1]{fontenc}    
\usepackage{hyperref}       
\usepackage{url}            
\usepackage{booktabs}       
\usepackage{amsfonts}       
\usepackage{nicefrac}       
\usepackage{microtype}      
\usepackage{lipsum}
\usepackage{graphicx}
\usepackage{pgfplots}
\usepackage{float}
\usepackage{amsmath}
\usepackage{adjustbox} 
\usepackage{subcaption}
\graphicspath{ {./images/} }

\title{Predict future sale}

\author{
 Ke Xue \\
 School of Cyberspace Science and Technology\\
  Beijing Institute of Technology\\
  Beijing, 100081 \\
  \texttt{3220231773@bit.edu.cn} \\
   \And
 Rongfei Fan \\
  School of Cyberspace Science and Technology\\
  Beijing Institute of Technology\\
  Beijing, 100081 \\
  \texttt{fanrongfei@bit.edu.cn} \\
  \And
 Shanping Yu \\
  School of Cyberspace Science and Technology\\
  Beijing Institute of Technology\\
  Beijing, 100081 \\
  \texttt{ysp@bit.edu.cn} \\
  \And
 Chang Sun \\
  School of Computer Science\\
  Beijing University of Posts and Telecommunications\\
  Beijing, 100876 \\
  \texttt{sunchang@bupt.edu.cn} \\
  \And
 Jianping An \\
  School of Cyberspace Science and Technology\\
  Beijing Institute of Technology\\
  Beijing, 100081 \\
  \texttt{an@bit.edu.cn} \\
}

\begin{document}

\title{DualStream Contextual Fusion Network: Efficient Target Speaker Extraction by Leveraging Mixture and Enrollment Interactions}

\maketitle

\begin{abstract}

Target speaker extraction focuses on extracting a target speech signal from an environment with multiple speakers by leveraging an enrollment. Existing methods predominantly rely on speaker embeddings obtained from the enrollment, potentially disregarding the contextual information and the internal interactions between the mixture and enrollment. In this paper, we propose a novel DualStream Contextual Fusion Network (DCF-Net) in the time-frequency (T-F) domain. 
Specifically, DualStream Fusion Block (DSFB) is introduced to obtain contextual information and capture the interactions between contextualized enrollment and mixture representation across both spatial and channel dimensions, and then rich and consistent representations are utilized to guide the extraction network for better extraction. 
Experimental results demonstrate that DCF-Net outperforms state-of-the-art (SOTA) methods, achieving a scale-invariant signal-to-distortion ratio improvement (SI-SDRi) of 21.6 dB on the benchmark dataset, and exhibits its robustness and effectiveness in both noise and reverberation scenarios. In addition, the wrong extraction results of our model, called \textit{target confusion problem}, reduce to 0.4\%, which highlights the potential of DCF-Net for practical applications.

\end{abstract}



\section{Introduction}

In real-life scenarios, cocktail party problems \cite{cherry1953some} are increasingly prevalent, which separates individual sound sources from a mixture of sounds accompanied with noise and reverberation.
Research \cite{conway2001cocktail,coch2005event,mesgarani2012selective}, indicates that humans, even infants as young as a few months old, can selectively focus on the content they wish to hear. Inspired by this capability, there is a strong interest in endowing machines with similar functionality.
Two fundamental and promising approaches have emerged. The first is speech separation (SS) \cite{wang2018supervised, luo2020dual,chen2020dual,wang2023tf,kalkhorani2024crossnet}, which aims to disentangle all speech signals from the mixture. 
However, speech separation is significantly limited by the requirement to know the number of audio sources in advance, which is often difficult to determine in practical scenarios, which hinders its practicality.

Another approach is target speaker extraction (TSE) \cite{zmolikova2023neural}, as a practical alternative solution for real-world applications. This method concentrates on isolating the speech of a particular target speaker from the mixture, rather than attempting to separate all speeches. Typically, this approach utilizes auxiliary information about the target speaker, i.e utilizing a fixed-dimensional speaker embedding processed by a speaker encoder with an {\it enrollment}
\cite{wang2018voicefilter,xu2019time, xu2020spex, ge2020spex+, xu2023adaptive, ge2021multi, wang2021neural,deng2020robust,yang2023target,liu2023x,peng2024target}. In these methods, the fixed-dimensional speaker embeddings are used to guide the main extraction network to extract the target speech. 

VoiceFilter \cite{wang2018voicefilter} is recognized as the pioneering model for TSE in {time-frequency (T-F)} domain. It consists of two primary components: a speaker encoder that generates discriminative {speaker embeddings} and a {T-F enhancement network} that utilizes {these embeddings} {to refine the T-F representations}. 
In \cite{vzmolikova2019speakerbeam}, scaled activations and factorized layer methods are creatively employed to boost the extraction performance. 
Nevertheless, these typically focus solely on magnitude information while {omitting}
the original phase of the mixed signal for reconstruction, which dramatically impacts the quality of the reconstructed signal.

To address the phase issue in T-F domain methods,
TseNet \cite{xu2019time} and the SpEx series \cite{xu2020spex, ge2020spex+, xu2023adaptive, ge2021multi, wang2021neural} 
{perform signal decomposition in the time domain and further employ multi-scale encoders to improve time-domain resolution, thereby enhancing extraction accuracy.} 
In \cite{deng2020robust, yang2023target, liu2023x, peng2024target} various methods on either enrollment or extraction network are exploited to achieve further enhancement in the time domain.
On the other hand, recently \cite{hao2024x} processes both the real and imaginary parts simultaneously in T-F domain, resolves phase issues and achieves superior results
.
By this means, it is hard to say whether T-F domain methods or time-domain methods dominate.
\begin{figure}[H]

\begin{minipage}[b]{1.0 \linewidth}
 \centering
 \hspace{3cm}  
 \centerline{\includegraphics[width= 0.7\columnwidth]{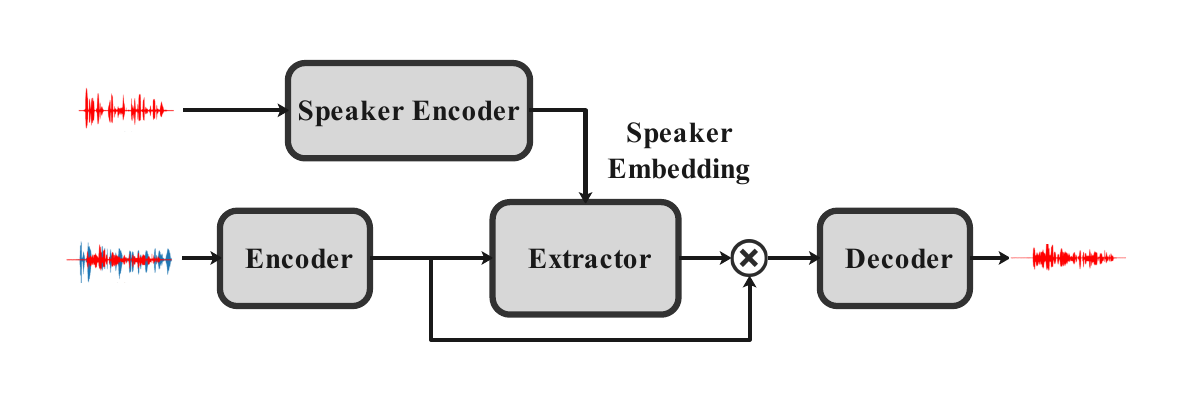}}
\end{minipage}

\caption{TSE network {leveraging} 
speaker embeddings.}

\label{f:fig1}

\end{figure}

All aforementioned approaches leverage speaker encoder to form speaker embedding from enrollment, as shown in Fig. \ref{f:fig1}. However, speaker encoder can only capture the unique voiceprint characteristics of the speaker from enrollment, disregards contextual information in the enrollment,which may overlook the features that could assist the extraction network in capturing more informative cues for separation.

Recently, several studies have explored alternative methods \cite{xiao2019single, zeng2023sef,xue2024target}.
To be specific, \cite{xiao2019single} and \cite{zeng2023sef} both separately processing the mixture signal and enrollment on time domain, \cite{xiao2019single} employs an attention mechanism to compute bias, while \cite{zeng2023sef} employs advanced conformer and then fuses the extracted feature sequences.
\cite{xue2024target}, as an alternative, establishes connections within the T-F domain at the audio level, jointly processing the mixture signal and enrollment within an interaction block.
Retrospecting the works in \cite{xiao2019single, zeng2023sef,xue2024target}, we identify the following limitations: separating processing of mixed signal and enrollment risks partial loss of target speaker information,  \cite{xiao2019single} relies solely on the magnitude spectrum. \cite{zeng2023sef} requires padding or truncation to align the frames between enrollment and mixed signals and involves complex multi-stage feature fusion across different modules. \cite{xue2024target} merely establishes connections within T-F domain at audio level and then simply concatenates the two components which may be insufficient.

To overcome the above shortage, we propose an advanced extraction network, named DualStream Contextual Fusion Network (DCF-Net), 
we adopt the complex spectrum in T-F domain instead of relying solely on the magnitude spectrum. Inspired by \cite{xue2024target}, we also use an interaction block to jointly process the mixture signal and enrollment and padding or truncation can be omitted. In addition, to address possible limitations of \cite{xue2024target}, we propose DualStream Fusion Block (DSFB),in which the MGI mechanism and the Squeeze-and-Excitation (SE) block are integrated to explore the contextual and internal information between the two components at the feature map level after multi-scale convolution.
The contributions of our work are as follows:

$\bullet$ We propose a target speaker extraction network called DCF-Net, the core of which is DualStream Fusion Block (DSFB), to retrieve richer contextual and internal information.

$\bullet$ We integrate the MGI mechanism and the Squeeze-and-Excitation (SE) block into the DualStream Fusion Block (DSFB) to enhance the audio information through adaptive recalibration of channel-wise feature responses.

$\bullet$ Extensive experiments on three datasets demonstrate the robustness and effectiveness of our approach in varied scenarios. Furthermore, the low incidence of the target confusion problem underscores its potential for practical applications.

\begin{figure*}[htb]

\begin{minipage}[b]{1.0 \linewidth}
 \centering
 \hspace{3cm}  
 \centerline{\includegraphics[width= \textwidth]{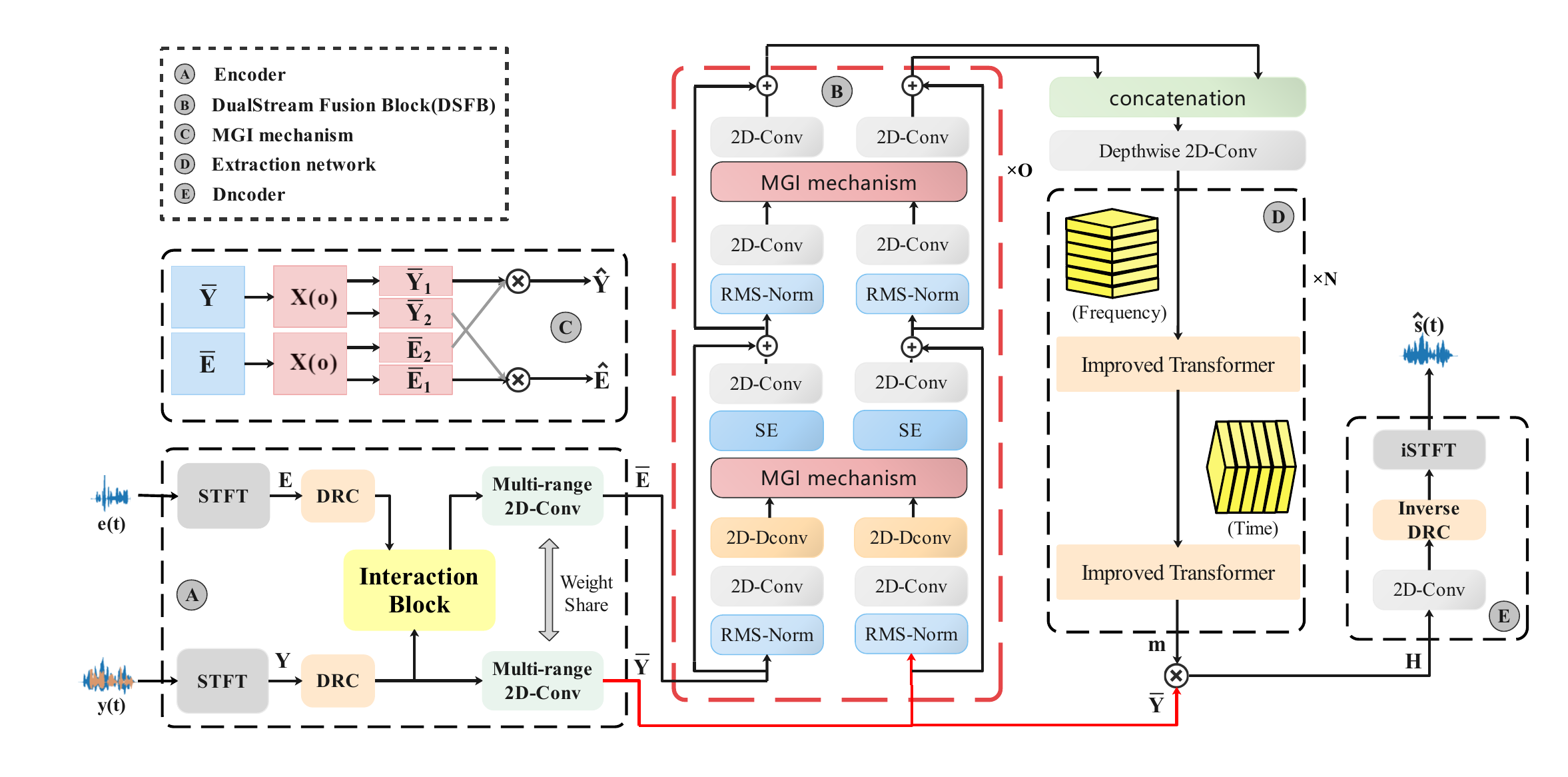}}
\end{minipage}

\caption{The overall architecture of our proposed network. There are four main components, different capital letters represent different parts of our model.}

\label{f:fig2}

\end{figure*}

\section{{Physical Model and Objective}}

Given the $y(t)$ and $e(t)$, where $y(t)$ represents the mixed signal and the $e(t)$ represents the enrollment of $s(t)$'s speaker, we have 
\begin{align}
    y(t) = s(t) + \sum_{i = 1}^{U} b_i(t), t = 1, \cdots, T
\end{align}
where $U$ is the number of interference sources, $b_i(t)$ is interference speech or background noise and reverberation for $i$th source, and $t$ is the time index of samplings ranging from 1 to $T$. 
Our purpose is to extract $\hat{s}(t)$ from $y(t)$ as an estimate of $s(t)$ referring to $e(t)$.

\section{DCF-Net}
\label{ssec:subhead}
The proposed model, as illustrated in Fig.\ref{f:fig2}, consists of four main components: encoder: Fig.\ref{f:fig2}-A, DualStream Fusion Block: Fig.\ref{f:fig2}-B, extraction network: Fig.\ref{f:fig2}-D, and decoder: Fig.\ref{f:fig2}-E. DualStream Fusion Block is the core of our network.
\subsection{Encoder}
\label{sssec:subhead}

\begin{figure}[H]

\begin{minipage}[b]{1.0 \linewidth}
 \centering
 \hspace{3cm}  
 \centerline{\includegraphics[width= 0.7\columnwidth]{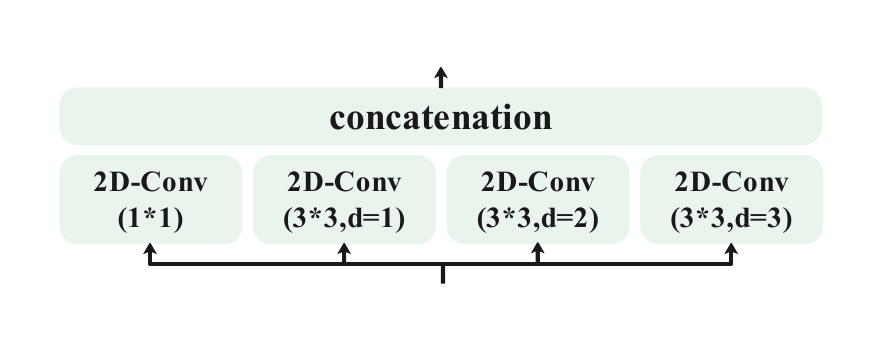}}
\end{minipage}

\caption{Multi-range 2D convolution.}

\label{f:fig0}

\end{figure}

The enrollment $ e (t) $ and the mixed signal $ y(t) $  are served as the input of the encoder, then transform into T-F representations $ E\in \mathbb{R}^{C \times F_1\times T_1}$ and $ Y\in \mathbb{R}^{C \times F \times T}$ via STFT, respectively, where $ C $ is the feature dimension, $ F_1 $ and $ F $ are the number of the frequency bins, $ T_1 $ and $ T $ are the frame number of signals. Subsequently, dynamic range compression (DRC) \cite{li2021importance} is applied to stabilize the amplitude variations. Following this, an interaction block \cite{xue2024target} is introduced, where the weighting matrices are first computed based on the spectrum of the enrollment and the mixture to obtain a contextualized enrollment spectrum to match the shape of the mixture spectrum. 
Finally, weighted-share multi-range 2D convolution is employed to generate the refined representations $\bar{E}\in \mathbb{R}^{C \times F \times T}$ and $\bar{Y}\in \mathbb{R}^{C \times F \times T}$, see Fig. \ref{f:fig0}.

\subsection{DualStream Fusion Block (DSFB)}
\label{sssec:subhead}
DSFB consists of two fully symmetric {channel flows}, receiving the outputs $ Y $ and $ \bar{E} $ of the encoder respectively, followed by RMS layer normalization (RMS-Norm) \cite{zhang2019root}. Compared to LayerNorm, RMS-Norm effectively stabilizes the magnitude of layer activations, which ensures invariance to the rescaling of weights and datasets, thereby promoting more robust and consistent model performance. Then a $1\times1$ 2D convolution is multiplied to double the channels of $ Y $ and \(\bar{E}\), preventing the loss of original feature information due to the halving of channels in the MGI mechanism. Then $3\times3$ 2D depthwise convolution follows to further extract the corresponding features. Two flows of the dual-stream information are fed into the MGI mechanism module. With regard to the MGI mechanism \cite{hu2023single}, which is illustrated in Fig.\ref{f:fig1}-C, we fully exploit the fine interaction between \({Y}\) and \(\bar{E}\), these operations can be formulated as follows:
\begin{alignat}{2}
    [{Y}_1, {Y}_2] &= X(L_2(L_1(\text{RMS-Norm}(Y)))) \\
    [\Bar{E}_1, \Bar{E}_2] &= X(L_2(L_1(\text{RMS-Norm}(\Bar{E})))) \\
    \hat{Y} = {Y}_1 \circ \Bar{E}_2,
    & \quad \hat{E} = {Y}_2 \circ \Bar{E}_1,
\end{alignat}
where $ L_1 $, $ L_2 $ represent $1\times1$ 2D convolution and $3\times3$ 2D depthwise convolution, respectively, $\hat{Y}$ and $\hat{E}$ are the outputs of the MGI mechanism, \(\circ\) represents element-wise multiplication, and $X(\cdot)$ is the function to be trained that separates the input signal into two output signals with equal number of channels. 
Clearly, the output $\hat{Y}$ and \(\hat{E}\) have their channel numbers halved after the MGI mechanism. 






Subsequently, squeeze-and-Excitation (SE) block \cite{hu2018squeeze} is integrated into DSFB to enhance its ability to distinguish the importance of different channels. The specific implementation of SE is as follows: Global average pooling is applied to the features, followed by a convolution to obtain a weight vector, which is conducted element-wise multiplication with the original feature to generate a weighted feature map. Through this way, each channel’s features are weighted according to their importance, which enhances the network’s focus on critical mixture and enrollment information. A $1\times 1$ 2D convolution is then followed. Finally, residual connections are applied, followed by a repetition of the previous operations, excluding the $3 \times 3$ 2D depthwise convolution and SE block in order to reduce additional computation. 
After $ O $ DSFB, the two output streams are concatenated, followed by a 2D convolution and { ReLU activation function}.

\subsection{Extraction Network}
\label{sssec:subhead}
As shown in Fig.\ref{f:fig1}-D, we employ a dual-path improved transformer proposed by \cite{chen2020dual} which is composed by Multi-Head Attention mechanism (MHA) and Bi-directional Long Short-Term Memory (BLSTM) .Compared with recurrent neural network (RNN), transformers are capable of handling and modeling long time series or sequences, avoiding the problem of limited receptive fields. And dual-path network \cite{luo2020dual} is an effective method to deal with tens of thousands of input sequences in speech extraction. More details can see \cite{chen2020dual}. This extraction network is operated to estimate a mask $m$ of the target speaker.

\subsection{Decoder}
\label{sssec:subhead}

As for decoder, the estimated feature tensor $H$ is realized by element-wise multiplying the mixture speech embedding (T-F representation) with the mask $m$, followed by a 2D convolution to transform back into target speaker’s T-F representation. Then the inverse DRC and the inverse STFT are employed to reconstruct the target speaker’s speech $\hat{s}(t)$. 

During training, scale-invariant signal-to-distortion ratio (SI-SDR) \cite{le2019sdr} is used as our loss function. Its role is to minimize the reconstruction loss of the signal while ignoring the signal's energy magnitude, which is defined as:

\begin{alignat}{3}
    \text{SI-SDR}(s, \hat{s}) = 10 \log_{10} \left( \frac{\| \tilde{s} \|^2}{\| \tilde{e} \|^2} \right)
\end{alignat}
where
\begin{alignat}{4}
        \tilde{s} &= \frac{\langle s, \hat{s} \rangle}{\| s \|^2} s, & \quad \tilde{e} &= \hat{s} - \tilde{s}.
\end{alignat}
here, $\tilde{s}$, $\tilde{e}$ represent clean and noise signals respectively, $\langle s, \hat{s} \rangle$ denotes the inner product of $s$ and $\hat{s}$, and $\tilde{s}$, $\tilde{s}$ are normalized to zero-mean prior to the calculation.


\section{Experimental Setup}
\label{sec:page-setup}
\subsection{Datasets}
\label{ssec:subhead}

\noindent $\bullet$ \textbf{WSJ0-2Mix} This dataset is based on the WSJ0 corpus \cite{garofolo1993csr} at a sampling rate 8kHz. 
Specifically, two speakers' utterances were randomly selected {from the WSJ0 ``si\_tr\_s'' corpus} to generate training and validation sets with a relative signal-to-noise (SNR) ratio between 0 and 5 dB. The reference for the target speaker is randomly selected and different from the mixed speech. Similarly, the test set was generated by randomly mixing two speakers' utterances from the WSJ0 ``si\_dt\_05'' and ``si\_et\_05'' corpora. 

\noindent $\bullet$ \textbf{WHAM!} \cite{wichern2019wham} This dataset consists of simulated mixtures of clean speech signals from the WSJ0 corpus, combined with various types of background noise at different signal-to-noise ratios. 

\noindent $\bullet$ \textbf{WHAMR!} \cite{maciejewski2020whamr} This dataset extends the WHAM! dataset by incorporating extra reverberation.

\subsection{Experimental Configurations}
\label{ssec:subhead}
In this paper, the channel dimension \(C\) is set to 256, and the number $F$ of frequency bins is 129. The number of attention heads is 4. The number of DSFB \(O\) is set as 2 and the \(N\) in the separation blocks are 6, respectively. The dimension fed into the transformer is 64 where the hidden dimension is 128. 
We trained the model for 150 epochs using the AdamW optimizer. For the learning rate schedule, we employed a WarmupCosineScheduler. Specifically, we implemented a linear warm-up phase with an initial learning rate of 0.0005 for the first 5 epochs, followed by cosine annealing to set the learning rate of the rest epochs.
\section{Results and Discussions}
\label{sec:page-dis}

In this section, we will compare the performance of our proposed model with existing models over WSJ0-2Mix, WHAM! and WHAMR! in signal reconstructing quality and target confusion problem, and perform some ablation experiments to verify the feasibility of our model configs. The performance metrics include Signal-to-Distortion Ratio improvement (SDRi), SI-SDR improvement (SI-SDRi). For each metric, higher values indicate better performance.

\begin{table}[H]
\centering
\caption{Comparison methods on WSJ0-2Mix dataset.}
\label{tab:sota}
\begin{adjustbox}{max width=0.8\columnwidth} 
\resizebox{\linewidth}{!}{
\begin{tabular}{|c|c|c|c|c|}
\hline
\textbf{Methods} & \textbf{Domian} & \textbf{Params(M)} & \textbf{SI-SDRi} & \textbf{SDRi}  \\ \hline
TseNet \cite{xu2019time} & Time & 9.0M & 12.2 & 12.8  \\ 
SpEx \cite{xu2020spex} & Time & 10.8M & 14.6 & 15.1  \\ 
SpEx+ \cite{ge2020spex+} & Time & 11.1M  & 15.7 & 15.9 \\ 
Adaptive-SpEx \cite{xu2023adaptive} & Time & - & 18.8 & 19.2  \\

SpEx++ \cite{ge2021multi} & Time & - & 18.0 & 18.4  \\ 
SpExsc \cite{wang2021neural} & Time & 28.4M & 19.0 & 19.2  \\
DPRNN-Spe-IRA \cite{deng2020robust} & Time & 2.9M & 17.5 & 17.7  \\ 
VEVEN \cite{yang2023target} & Time & 2.6M & 19.0 & 19.2  \\ 
X-SepFormer \cite{liu2023x} & Time & - & 19.1 & 19.7  \\ \hline

SEF-Net \cite{zeng2023sef} & Time & 27M & 17.2 & 17.6  \\
CIENet-mDPTNet \cite{xue2024target} & T-F & 2.9M & 21.4 & 21.6  \\ \hline
DCF-Net & T-F & \textbf{3.9M} & \textbf{21.6} & \textbf{21.7}  \\ \hline
\end{tabular}
}
\end{adjustbox}
\end{table}

\subsection{Comparison SOTA methods on WSJ0-2Mix dataset}
\label{ssec:subhead}

Table \ref{tab:sota} present the comparative results of our proposed DCF-Net and other representative models on WSJ0-2Mix. To be specific, there are two categories of benchmark methods. The first category, from TseNet to X-SepFormer, fundamentally relies on speaker embeddings to extract target speech. For the second category, like CIE-mDPTNet, exploits the contextual information contained in the enrollment instead of speaker embeddings and achieve better results, therefore based on this, we use DSFB to better capture the interactions between contextualized enrollment and mixture representation. Furthermore, our proposed model DCF-Net, demonstrates 21.9dB and 22.1dB on SI-SDRi and SDRi, respectively.


\subsection{Comparison methods on more complicated dataset }
\label{ssec:subhead}

In this subsection, our proposed models are compared with several methods on WHAM! and WHAMR! dataset.

\begin{table}[h]
\centering
\caption{Comparison methods on WHAM! and WHAMR!}
\label{tab:sota2}
\begin{adjustbox}{max width=0.8\columnwidth} 
\resizebox{\linewidth}{!}{
\begin{tabular}{|c|c|c|c|c|}
\hline
\textbf{Methods} & \multicolumn{2}{c|}{\textbf{WHAM!}} & \multicolumn{2}{c|}{\textbf{WHAMR!}} \\ \cline{2-5}
                 & \textbf{SI-SDRi} & \textbf{SDRi} & \textbf{SI-SDRi} & \textbf{SDRi} \\ \hline
SpEx \cite{xu2020spex} & 12.2 & 13.0 & 10.3 & 9.5 \\
SpEx+ \cite{ge2020spex+} & 13.1 & 13.6 & 10.9 & 10.0 \\
SpEx++ \cite{ge2021multi} & 14.3 & 14.7 & 11.7 & 10.7 \\
DPRNN-Spe-IRA \cite{deng2020robust} & 14.2 & 14.6 & - & - \\
CIENet-mDPTNet \cite{xue2024target} & 16.6 & 17.0 & 15.7 & 14.3 \\ \hline
DCF-Net & \textbf{16.8} & \textbf{17.3} & \textbf{15.8} & \textbf{14.5} \\ \hline
\end{tabular}}
\end{adjustbox}
\end{table}

As shown in Table \ref{tab:sota2}, all models showcase significant performance degradation in the presence of noise and reverberation. However, inspite of this, our models still achieve higher performance in both noise and reverberation scenarios, which exhibit the robustness of our model.


\subsection{Target confusion problem on WSJ0-2Mix}

\begin{figure}[htbp]  
    \centering
    \begin{subfigure}{0.49\columnwidth}  
        \centering
        \includegraphics[width=\linewidth]{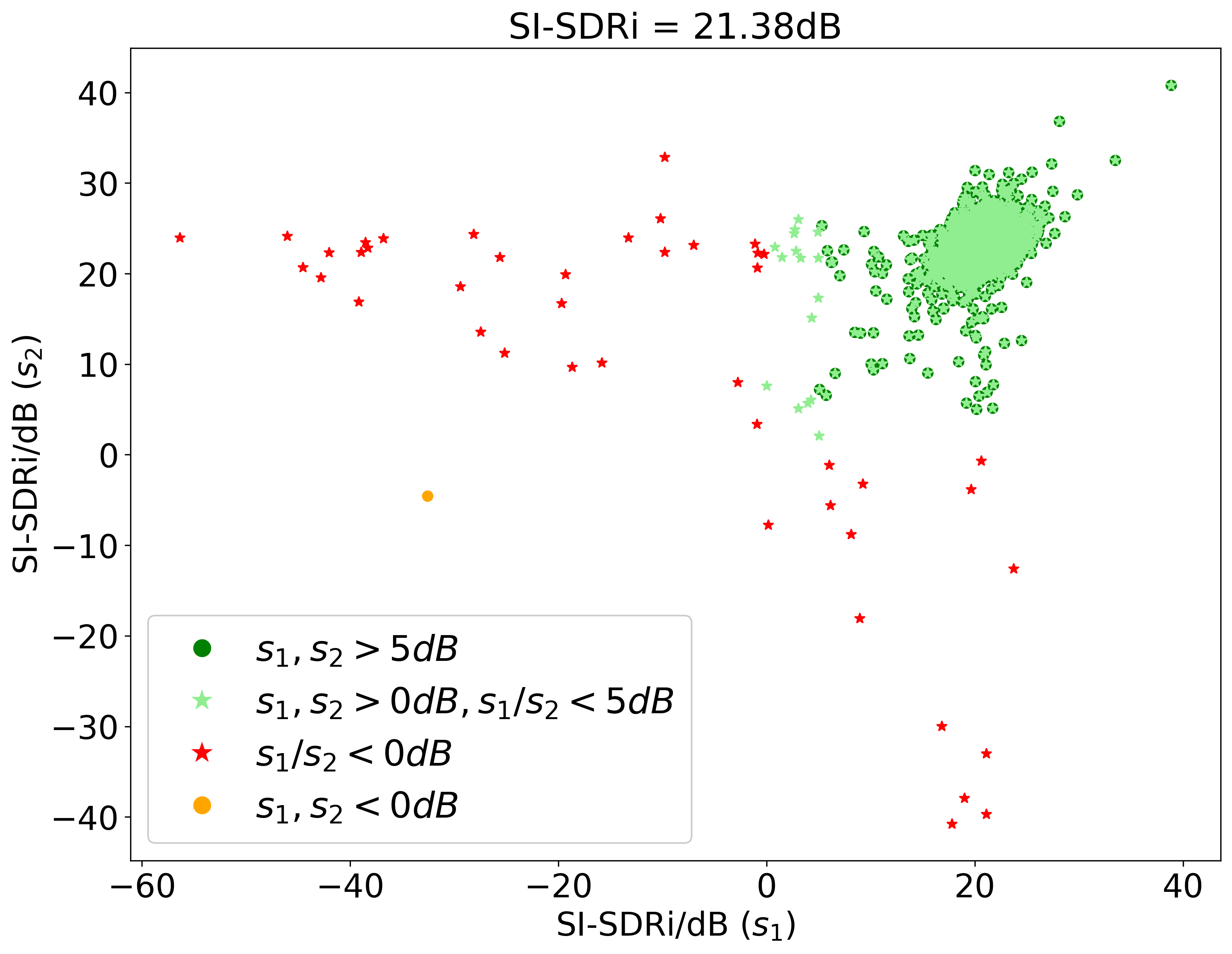}  
        \caption{(TCP) in CIENet($1\%$).}  
        \label{fig:scatter_plot}
    \end{subfigure}
    \hfill  
    \begin{subfigure}{0.49\columnwidth}  
        \centering
        \includegraphics[width=\linewidth]{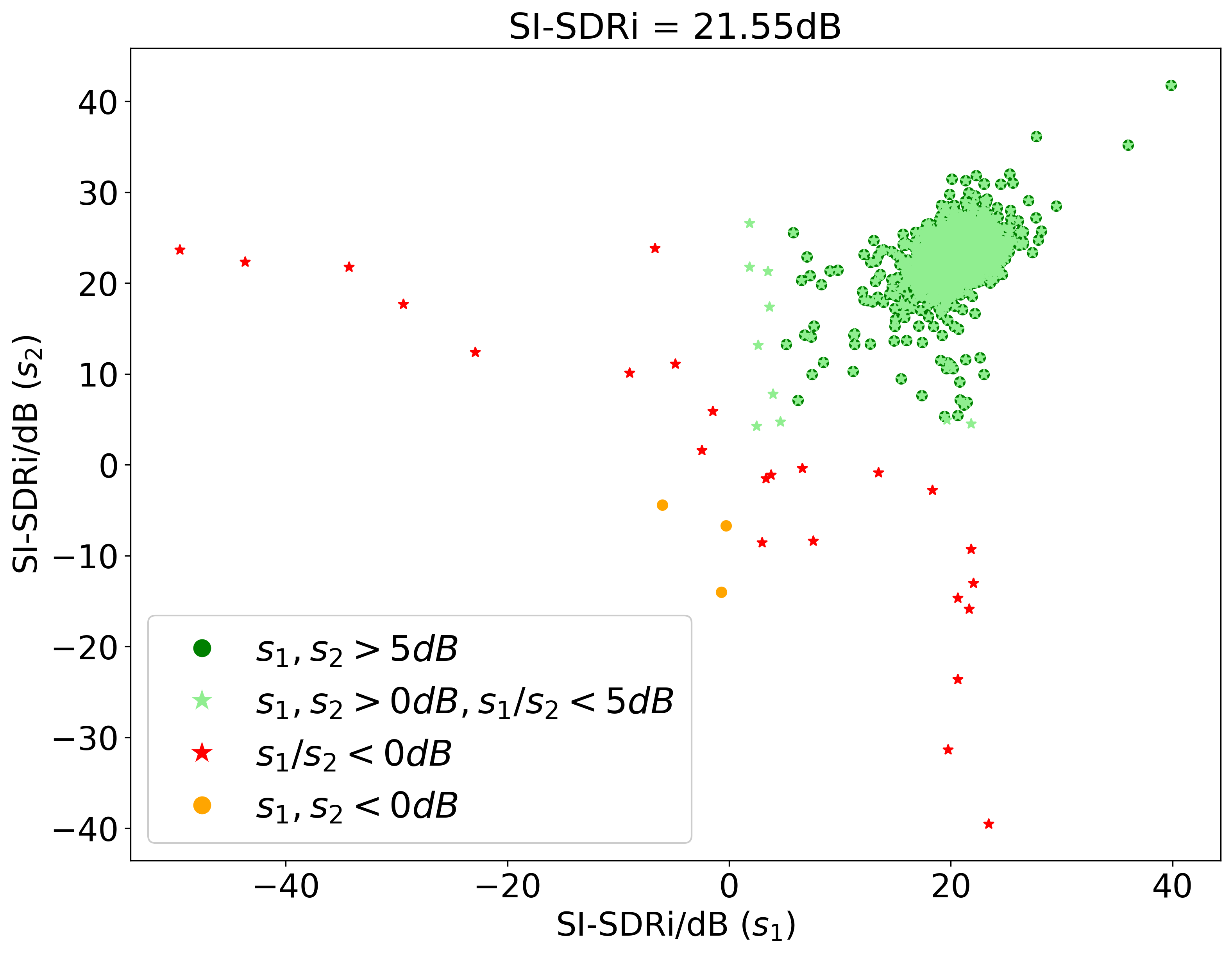}  
        \caption{(TCP) in DCF-Net($0.4\%$).}  
        \label{fig:scatter_plot_cienet}
    \end{subfigure}
    
    \caption{The target confusion problem (TCP) in CIENet $(a)$ and DCF-Net $(b)$. These two are the distribution in terms of SI-SDRi on WSJ0-2Mix test set, each axis corresponds to a speaker. $s1$, $s2$ $<$ $0dB$ denotes that SI-SDRi of both $s1$ and $s2$ are below $0dB$, while $s1/s2 < 0dB$ means that either of them is below $0dB$, both of two mean the target confusion problem happens.}  
    \label{fig:comparison_scatter_plots}  
\end{figure}

We use the same scatter diagram deployed in \cite{zhao2022target} to compare the target confusion problem (TCP) As shown in Fig. \ref{fig:comparison_scatter_plots}. Red and orange indicate that the model has encountered target confusion problem (TCP), meaning it has extracted incorrect speech. Compared with CIENet (Fig. \ref{fig:scatter_plot}) , our model (Fig. \ref{fig:scatter_plot_cienet}) has lower rate ($0.4\%$) . The introduction of the DSFB module might have allowed the model to optimally utilize the information from the enrollment, thus decreasing the likelihood of extracting error rate.
\label{ssec:subhead}

\subsection{Ablation  experiments}
\label{ssec:subhead}

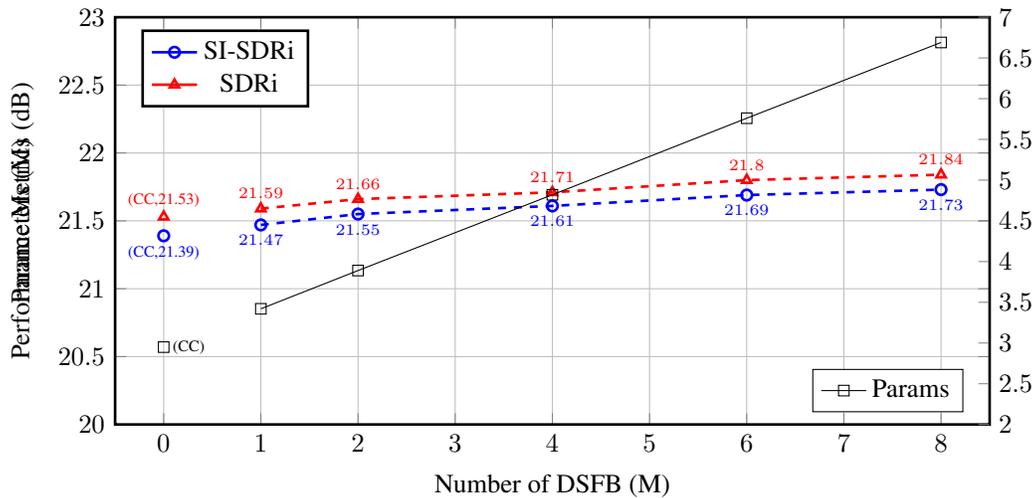
\begin{figure}[htbp]
    \centering
    \begin{tikzpicture}
        \begin{axis}[
            xlabel={Number of DSFB (M)},
            ylabel={Performance Metrics (dB)},
            xmin=-0.5, xmax=8.5,
            ymin=20, ymax=23,
            xtick={0,1,...,8},
            ytick={20,20.5,...,23},
            grid=both, 
            width=0.8\columnwidth,
            height=7cm,
            legend pos=north west,
            ymajorgrids=true, 
            xmajorgrids=true,
            line width=1pt,
            mark size=2pt,
            every axis plot post/.append style={mark options={solid}},
            ]
            
            \addplot[
                color=blue,
                mark=o,
                ]
                coordinates {
                (0,21.39)
                } node[anchor=north, pos=1, font=\tiny] {(CC,21.39)};
            \addplot[
                color=red,
                mark=triangle,
                ]
                coordinates {
                (0,21.53)
                } node[anchor=south, pos=1, font=\tiny] {(CC,21.53)}; 
                
            \addplot[
                color=blue,
                dashed,
                mark=o,
                nodes near coords, 
                every node near coord/.append style={
                    font=\tiny,
                    /pgf/number format/.cd, fixed, precision=2,
                },
                 nodes near coords align={below}, 
                ]
                coordinates {
                (1,21.47)(2,21.55)(4,21.61)(6,21.69)(8,21.73)
                };
                
                \addlegendentry{SI-SDRi}
                
            \addplot[
                color=red,
                dashed,
                mark=triangle,
                nodes near coords, 
                every node near coord/.append style={
                    font=\tiny,
                    /pgf/number format/.cd, fixed, precision=2,
                },
                nodes near coords align={above}, 
                ]
                coordinates {
                (1,21.59)(2,21.66)(4,21.71)(6,21.80)(8,21.84)
                };
                \addlegendentry{SDRi}
                
        \end{axis}
        
        \begin{axis}[
            axis y line*=right,
            axis x line=none, 
            ylabel={Parameters (M)},
            xmin=-0.5, xmax=8.5, 
            ymin=2, ymax=7,
            ytick={2,2.5,...,7},
            width=0.8\columnwidth,
            height=7cm,
            legend pos=south east,
            ]
            
            \addplot[
                color=black,
                mark=none,
                mark=square,
                ]
                coordinates {
                (1,3.42)(2,3.89)(4,4.82)(6,5.76)(8,6.69)
                };
                \addlegendentry{Params}
                
            \addplot[
                color=black,
                mark=square,
                ]
                coordinates {
                (0,2.95)
                }node[anchor=west, pos=1, font=\tiny] {(CC)};
                
        \end{axis}
    \end{tikzpicture}
    \caption{Performance comparison of varying numbers of DSFB (O) on the WSJ0-2Mix dataset, including parameter(M) count.}
    \label{fig:performance_comparison_with_params}
\end{figure}

This subsection shows model performance based on different number of DSFB. $O$ is the number of DSFB, from 0 to 8 as shown in Fig.\ref{fig:performance_comparison_with_params}. When $O$ is 0, we demonstrate Concatenation Connection (CC) to the output of the encoder. Obviously, the methods with no DSFB have Sub-optimal performance, which conversely shows the effectiveness of the DSFB. With the increase number of $O$, the larger of $O$, the smaller improvement of SI-SDRi and SDRi. In order to compromise these two, we choose $ O $ is 2.


\begin{table}[h]
\centering
\caption{Comparison model with diferent network}
\label{tab:sota4}
\begin{adjustbox}{max width=0.8\columnwidth} 
\resizebox{\linewidth}{!}{
\begin{tabular}{|c|c|c|c|c|}
\hline
\textbf{Methods} & \textbf{Params(M)} & \textbf{SI-SDRi} & \textbf{SDRi} \\ \hline
DFC-Net(RNN) & 3.7M & 20.8 & 21.0  \\ 
DFC-Net(BT) & 3.8M & 21.3 & 21.5  \\ 
DFC-Net(IT) & 3.9M  & \textbf{21.6} & \textbf{21.7}  \\ \hline

\end{tabular}
}
\end{adjustbox}
\end{table}

Table \ref{tab:sota4} presents the different extraction network, DFC-Net (RNN) means we replace dual path trasnformer with dual path RNN. The following two are employed with base transformer (BT) \cite{vaswani2017attention} and improved transformer (IT) \cite{chen2020dual}. As we can see, IT yields the most effective performance.

\section{Conclusion}

In this work, we propose a T-F domain target speaker separation network DCF-Net, in which DualStream Fusion Block is introduced to capture the interactions between contextualized enrollment and mixture representation across both spatial and channel dimensions. Numerical results show DCF-Net could outperform existing methods over the benchmark datasets. Future work we will explore more effective channel attention instead of SE in DSFB, and design more useful extraction network.

\newpage

\bibliography{article,refs}

\begin{thebibliography}{10}
\providecommand{\url}[1]{#1}
\csname url@samestyle\endcsname
\providecommand{\newblock}{\relax}
\providecommand{\bibinfo}[2]{#2}
\providecommand{\BIBentrySTDinterwordspacing}{\spaceskip=0pt\relax}
\providecommand{\BIBentryALTinterwordstretchfactor}{4}
\providecommand{\BIBentryALTinterwordspacing}{\spaceskip=\fontdimen2\font plus
\BIBentryALTinterwordstretchfactor\fontdimen3\font minus
  \fontdimen4\font\relax}
\providecommand{\BIBforeignlanguage}[2]{{%
\expandafter\ifx\csname l@#1\endcsname\relax
\typeout{** WARNING: IEEEtran.bst: No hyphenation pattern has been}%
\typeout{** loaded for the language `#1'. Using the pattern for}%
\typeout{** the default language instead.}%
\else
\language=\csname l@#1\endcsname
\fi
#2}}
\providecommand{\BIBdecl}{\relax}
\BIBdecl

\bibitem{cherry1953some}
E.~C. Cherry, ``Some experiments on the recognition of speech, with one and
  with two ears,'' \emph{J. Acoust. Soc. Am.}, vol.~25, no.~5, pp. 975--979,
  1953.

\bibitem{conway2001cocktail}
A.~R. Conway, N.~Cowan, and M.~F. Bunting, ``The cocktail party phenomenon
  revisited: The importance of working memory capacity,'' \emph{Psychon. Bull.
  Rev.}, vol.~8, pp. 331--335, 2001.

\bibitem{coch2005event}
D.~Coch, L.~D. Sanders, and H.~J. Neville, ``An event-related potential study
  of selective auditory attention in children and adults,'' \emph{J. Cogn.
  Neurosci.}, vol.~17, no.~4, pp. 605--622, 2005.

\bibitem{mesgarani2012selective}
N.~Mesgarani and E.~F. Chang, ``Selective cortical representation of attended
  speaker in multi-talker speech perception,'' \emph{Nature}, vol. 485, no.
  7397, pp. 233--236, 2012.

\bibitem{wang2018supervised}
D.~Wang and J.~Chen, ``Supervised speech separation based on deep learning: An
  overview,'' \emph{IEEE/ACM Trans. Audio Speech Lang. Process.}, vol.~26,
  no.~10, pp. 1702--1726, 2018.

\bibitem{luo2020dual}
Y.~Luo, Z.~Chen, and T.~Yoshioka, ``Dual-path rnn: efficient long sequence
  modeling for time-domain single-channel speech separation,'' in \emph{in
  Proc. IEEE ICASSP}, 2020, pp. 46--50.

\bibitem{chen2020dual}
J.~Chen, Q.~Mao, and D.~Liu, ``Dual-path transformer network: Direct
  context-aware modeling for end-to-end monaural speech separation,''
  \emph{arXiv preprint arXiv:2007.13975}, 2020.

\bibitem{wang2023tf}
Z.-Q. Wang, C.~Samuele, and C.~Shukjae, ``Tf-gridnet: Making time-frequency
  domain models great again for monaural speaker separation,'' in \emph{in
  Proc. IEEE ICASSP}, 2023, pp. 1--5.

\bibitem{kalkhorani2024crossnet}
K.~V.~Ahmadi and D.~Wang, ``Crossnet: Leveraging global, cross-band,
  narrow-band, and positional encoding for single-and multi-channel speaker
  separation,'' \emph{arXiv preprint arXiv:2403.03411}, 2024.

\bibitem{zmolikova2023neural}
Z.~Katerina, D.~Marc, and O.~Tsubasa, ``Neural target speech extraction: An
  overview,'' \emph{IEEE Signal Process. Mag.}, vol.~40, no.~3, pp. 8--29,
  2023.

\bibitem{wang2018voicefilter}
Q.~Wang, H.~Muckenhirn, and K.~Wilson, ``Voicefilter: Targeted voice separation
  by speaker-conditioned spectrogram masking,'' \emph{arXiv preprint
  arXiv:1810.04826}, 2018.

\bibitem{xu2019time}
C.~Xu, W.~Rao, E.~Chng, and H.~Li, ``Time-domain speaker extraction network,''
  in \emph{2019 IEEE Automatic Speech Recognition and Understanding Workshop
  (ASRU)}.\hskip 1em plus 0.5em minus 0.4em\relax IEEE, 2019, pp. 327--334.

\bibitem{xu2020spex}
------, ``Spex: Multi-scale time domain speaker extraction network,''
  \emph{IEEE/ACM Trans. Audio Speech Lang. Process.}, vol.~28, pp. 1370--1384,
  2020.

\bibitem{ge2020spex+}
M.~Ge, C.~Xu, L.~Wang, E.~Chng, J.~Dang, and H.~Li, ``Spex+: A complete time
  domain speaker extraction network,'' \emph{arXiv preprint arXiv:2005.04686},
  2020.

\bibitem{xu2023adaptive}
X.~Xu, D.~Yan, and D.~Li, ``Adaptive-spex: Local and global perceptual modeling
  with speaker adaptation for target speaker extraction,'' in \emph{2023 IEEE
  International Conference on Systems, Man, and Cybernetics (SMC)}.\hskip 1em
  plus 0.5em minus 0.4em\relax IEEE, 2023, pp. 342--347.

\bibitem{ge2021multi}
M.~Ge, C.~Xu, L.~Wang, E.~S. Chng, J.~Dang, and H.~Li, ``Multi-stage speaker
  extraction with utterance and frame-level reference signals,'' in \emph{in
  Proc. IEEE ICASSP}, 2021, pp. 6109--6113.

\bibitem{wang2021neural}
W.~Wang, C.~Xu, M.~Ge, and H.~Li, ``Neural speaker extraction with
  speaker-speech cross-attention network.'' in \emph{in Proc. Interspeech},
  2021, pp. 3535--3539.

\bibitem{deng2020robust}
C.~Deng, S.~Ma, Y.~Zhang, Y.~Sha, H.~Zhang, H.~Song, and X.~Li, ``Robust
  speaker extraction network based on iterative refined adaptation,''
  \emph{arXiv preprint arXiv:2011.02102}, 2020.

\bibitem{yang2023target}
L.~Yang, W.~Liu, L.~Tan, Y.~Jaemo, and M.~Han-Gil, ``Target speaker extraction
  with ultra-short reference speech by ve-ve framework,'' in \emph{in Proc.
  IEEE ICASSP}, 2023, pp. 1--5.

\bibitem{liu2023x}
K.~Liu, Z.~Du, X.~Wan, and H.~Zhou, ``X-sepformer: End-to-end speaker
  extraction network with explicit optimization on speaker confusion,'' in
  \emph{in Proc. IEEE ICASSP}, 2023, pp. 1--5.

\bibitem{peng2024target}
J.~Peng, D.~Marc, and O.~Tsubasa, ``Target speech extraction with pre-trained
  self-supervised learning models,'' in \emph{in Proc. IEEE ICASSP}, 2024, pp.
  10\,421--10\,425.

\bibitem{vzmolikova2019speakerbeam}
K.~{\v{Z}}mol, M.~Delcroix, K.~Kinoshita, T.~Ochiai, T.~Nakatani, and L.~J.
  Burget, ``Speakerbeam: Speaker aware neural network for target speaker
  extraction in speech mixtures,'' \emph{IEEE J. Sel. Top. Signal Process.},
  vol.~13, no.~4, pp. 800--814, 2019.

\bibitem{hao2024x}
F.~Hao, X.~Li, and C.~Zheng, ``X-tf-gridnet: A time--frequency domain target
  speaker extraction network with adaptive speaker embedding fusion,''
  \emph{Information Fusion}, vol. 112, 2024.

\bibitem{xiao2019single}
X.~e.~a. Xiao, ``Single-channel speech extraction using speaker inventory and
  attention network,'' in \emph{in Proc. IEEE ICASSP}, 2019, pp. 86--90.

\bibitem{zeng2023sef}
B.~Zeng, H.~Suo, Y.~Wan, and M.~Li, ``Sef-net: Speaker embedding free target
  speaker extraction network,'' \emph{Proc. Interspeech}, pp. 3452--3456, 2023.

\bibitem{xue2024target}
X.~Yang, C.~Bao, J.~Zhou, and X.~Chen, ``Target speaker extraction by directly
  exploiting contextual information in the time-frequency domain,'' in \emph{in
  Proc. IEEE ICASSP}, 2024, pp. 10\,476--10\,480.

\bibitem{li2021importance}
A.~Li, C.~Zheng, R.~Peng, and X.~Li, ``On the importance of power compression
  and phase estimation in monaural speech dereverberation,'' \emph{J. Acoust.
  Soc. Am. Express Lett.}, vol.~1, no.~1, 2021.

\bibitem{zhang2019root}
B.~Zhang and R.~Sennrich, ``Root mean square layer normalization,'' \emph{Adv.
  Neural Inf. Process. Syst.}, vol.~32, 2019.

\bibitem{hu2023single}
Q.~Hu and X.~Guo, ``Single image reflection separation via component synergy,''
  in \emph{Proceedings of the IEEE/CVF International Conference on Computer
  Vision}, 2023, pp. 13\,138--13\,147.

\bibitem{hu2018squeeze}
J.~Hu, L.~Shen, and G.~Sun, ``Squeeze-and-excitation networks,'' in
  \emph{Proceedings of the IEEE conference on computer vision and pattern
  recognition}, 2018, pp. 7132--7141.

\bibitem{le2019sdr}
L.~Jonathan, W.~Scott, E.~Hakan, and H.~John, ``Sdr--half-baked or well done?''
  in \emph{in Proc. IEEE ICASSP}, 2019, pp. 626--630.

\bibitem{garofolo1993csr}
G.~John, P.~David, G.~Doug, and P.~David, ``Csr-i (wsj0) complete ldc93s6a,''
  \emph{Web Download. Philadelphia: Linguistic Data Consortium}, vol.~83, 1993.

\bibitem{wichern2019wham}
G.~Wichern, J.~Antognini, M.~Flynn, L.~R. Zhu, E.~McQuinn, and J.~L. Roux,
  ``Wham!: Extending speech separation to noisy environments,'' \emph{arXiv
  preprint arXiv:1907.01160}, 2019.

\bibitem{maciejewski2020whamr}
M.~Maciejewski, G.~Wichern, E.~McQuinn, and J.~Le~Roux, ``Whamr!: Noisy and
  reverberant single-channel speech separation,'' in \emph{in Proc. IEEE
  ICASSP}, 2020, pp. 696--700.

\bibitem{zhao2022target}
Z.~Zhao, D.~Yang, R.~Gu, H.~Zhang, and Y.~Zou, ``Target confusion in end-to-end
  speaker extraction: Analysis and approaches,'' \emph{arXiv preprint
  arXiv:2204.01355}, 2022.

\bibitem{vaswani2017attention}
A.~Vaswani, ``Attention is all you need,'' \emph{Adv. Neural Inf. Process.
  Syst.}, 2017.

\end{thebibliography}
\bibliographystyle{IEEEtran}

\end{document}